\documentclass[doublecol]{epl3}
\usepackage{graphicx}
\usepackage{epstopdf}
\usepackage{bm}% bold math
\usepackage{subfigure}
\usepackage{sidecap}
\usepackage{amsmath}
\usepackage{ltxgrid}
\usepackage{lipsum}
\newcommand{\be}{\begin{equation}}
\newcommand{\ee}{\end{equation}}

\newcommand{\bea}{\begin{eqnarray}}
\newcommand{\eea}{\end{eqnarray}}
\newcommand{\bd}{\begin{displaymath}}
\newcommand{\ed}{\end{displaymath}}
\newcommand{\ba}{\begin{array}}
\newcommand{\ea}{\end{array}}
\newcommand{\bi}{\begin{itemize}}
\newcommand{\ei}{\end{itemize}}
\newcommand{\bc}{\begin{center}}
\newcommand{\ec}{\end{center}}
\newcommand{\bfl}{\begin{flushleft}}
\newcommand{\efl}{\end{flushleft}}
\newcommand{\bfr}{\begin{flushright}}
\newcommand{\efr}{\end{flushright}}

\newcommand{\oi}{{\rm i}}

\newcommand{\downdown}{ |\!\! \downarrow\downarrow\rangle }
\newcommand{\downup}{ |\!\! \downarrow\uparrow\rangle }
\newcommand{\updown}{ |\!\! \uparrow\downarrow\rangle }
\newcommand{\upup}{ |\!\!\uparrow\uparrow\rangle }

\newcommand{\bl}{\begin{aligned}}
\newcommand{\el}{\end{aligned}}

\def\6{\partial}

\def\no{\nonumber \\}

\def\={\!\!\!&=&\!\!\!}
\def\+{\!\!\!&&\!\!\!+~}
\def\-{\!\!\!&&\!\!\!-~}

\title{Gapped Quantum Criticality  Gains Long Time Quantum Correlations}
\author{R. Jafari\inst{1,2}
\and Alireza Akbari\inst{1,3,4}
}
\shortauthor{R. Jafari and A. Akbari}

\institute{
$^{1}$Asia Pacific Center for Theoretical Physics (APCTP), Pohang, Gyeongbuk, 790-784, Korea\\
$^{2}$Department of Physics, Institute for Advanced
Studies in Basic Sciences (IASBS), Zanjan 45137-66731, Iran\\
$^{3}$Department of Physics, POSTECH, Pohang, Gyeongbuk 790-784, Korea\\
$^{4}$Max Planck POSTECH Center for Complex Phase Materials, POSTECH,
Pohang 790-784, Korea\\
  }
\pacs{03.65.Yz}{Decoherence; open systems; quantum statistical methods}
\pacs{03.65.Ta}{Foundations of quantum mechanics; measurement theory}
\pacs{05.70.Jk}{Critical point phenomena}
\pacs{03.67.Mn}{Entanglement measures, witnesses, and other characterizations}

\abstract{
We show  gapped critical environment could remarkably
prevent an enhanced decay of decoherence factor and quantum correlations at the critical point,
which is nontrivially different from the ones in a gapless critical environment (Quan, et.al Phys. Rev. Lett. \textbf{96}, 140604 (2006)). The  quantum correlations display  very fast decaying to their local  minimum  at the critical point   while maximum decaying occurs away from this point.
In particular, our results imply that  collapse of decoherence factor is not indicator of a quantum phase transition of environment as opposed to what happens in a gapless criticallity.
In the week coupling regime, the relaxation  time, at which the quantum correlations touch rapidly   local minima,  shows a power-law singularity as a function of gap.  Furthermore, quantum correlations decay exponentially with second power of relaxation time.
Our results
are important for a better understanding and characterisation of gap critical environment  and its ability as entanglers in open  quantum systems.
 }

\begin{document}

\maketitle
%==============================================================

\twocolumngrid
%%%%%%%%%%%%%%%%%%%%%%%%%%%%%%%%%%%%%%%%%%%%%%%%%%%%%%%%%%%%%%%%%%%%%
%\section{Introduction \label{introduction}}

Quantum correlations (QCs)  are of primary  importance
in quantum information \cite{Barenco1, Pereira} and quantum computation \cite{Barenco2, Grover,Jafari:2010aa}.
They are   related to the basic issue of
understanding the nature of non-locality in quantum mechanics \cite{Einstein, Bell}.
Quantum systems used in quantum information processing inevitably interact with the
surrounding environment.
These correlated  surrounding systems induce quantum decoherency which    plays a key role in the  understanding of the quantum
to  the classical transition \cite{Zurek1,Paz}.
As a result,
in the last decade  a lot of efforts have been devoted to
investigate QCs  dynamics and  decoherence factors of central systems
in various environments~\cite{Almeida,Cormick:2008aa}.
The decoherence of a system coupled to a spin environment with  quantum phase transition (QPT) has been  investigated  intensively in various studies~\cite{Yuan2,Cucchietti3,Mostame,Sun:2007aa,Cheng:2009aa,Cormick:2008aa,Liu:2009aa,You-W.-L.:2010aa,Hu:2010aa,Luo:2011aa}.
Quan et al.~\cite{Quan} considered induction of the Ising-type correlated environment on the Loschmidt echo (LE), and found that the decaying behavior of  LE is best enhanced by gapless QPT of the surrounding system.
Rossini et al.~\cite{Rossini} depicted that in the short time region the LE decays as a Gaussian. However for long time limits they found that  it approaches an asymptotic value,  which strongly depends on the strength of the transverse magnetic field.
Further,  the decoherence of a system coupled to a spin environment with QPT has been  investigated~\cite{Yuan2,Cucchietti3,Mostame}.
%

%%%%%%%%%%%%%%%%%%%%%%%  Fig.1   %%%%%%%%%%%%%%%%%%%%%%%
\begin{figure*}[t]
\centerline{\includegraphics[width=\linewidth]{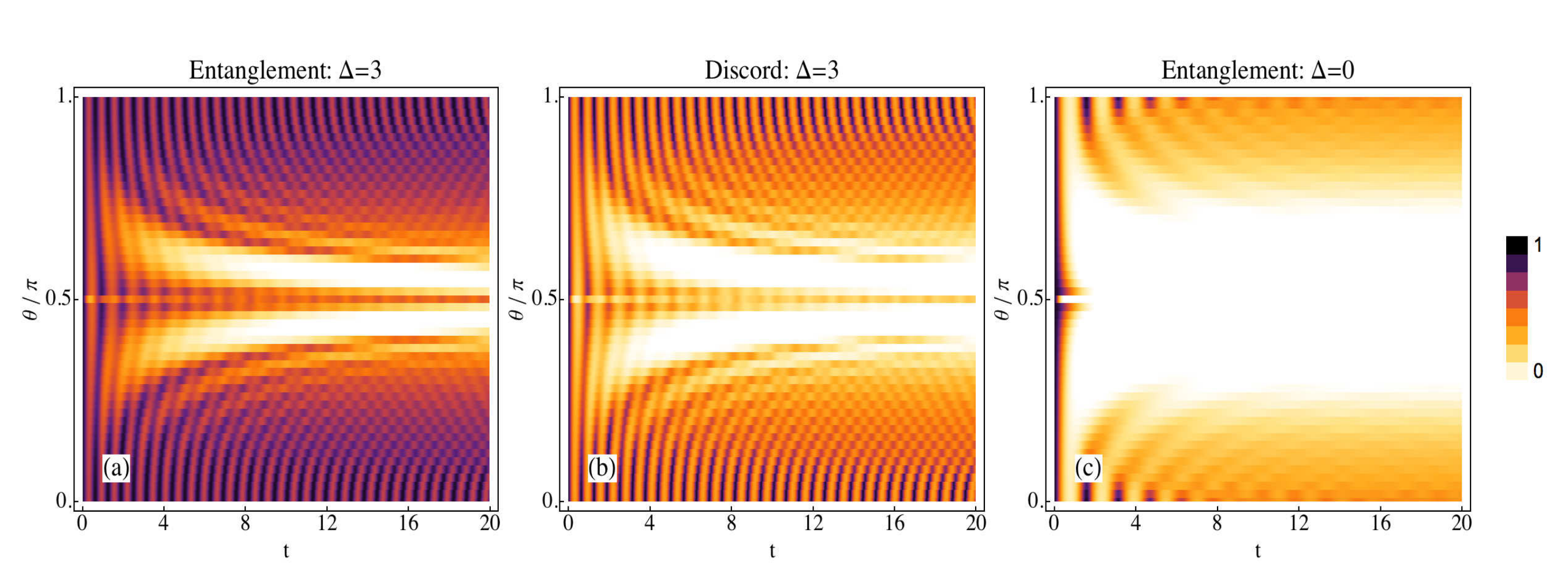}
}
\caption{(Color online) (a) Density plot of entanglement as a function of
time, for $J_{o}=1$, and $ J_{e}=4$. (b)  The variation of the quantum discord as function of time $t$ where the Hamiltonian parameters set as $J_{o}=1$, and $  J_{e}=4$.
(c) Entanglement evaluation for the gapless critical environment ($J_{o}=J_{e}=1$).
We consider  $N=400$ and weak-coupling
between qubits and spin bath $g=0.1$.
}
\label{fig1}
\end{figure*}
%%%%%%%%%%%%%%%%%%%%%%%%%%%%%%%%%%%%%%%%%%%%%%%%%%%%%%%%

The quantum phase transition occurs at a level crossing point (gapless phase transition) or converged point of avoided level crossing point (gapped critical point) \cite{Sachdev}. Because of the convergence of the energy levels at the critical point (CP),
some special dynamic features may appear in the dynamic evolution of the central  system in contact  with
an environment with QPT.
 It is shown that the disentanglement of the central  system is greatly enhanced
by the gapless quantum criticality  of the environment~\cite{Yuan2}. Furthermore, the decoherence induced by the critical environment may display some universal features~\cite{Cucchietti3}.
Since there exist separate states
showing non-classical behaviour  without entanglement~\cite{Horodecki, Niset, Datta}, the
quantum entanglement does not include all type of  quantum correlations and their non-classical properties.
Therefore, other measures of quantum correlations are expected.

A new but
promising notion is quantum discord first introduced by Ollivier and Zurek \cite{Ollivier},
which can effectively capture all QCs in a quantum system.
Recently,   the dynamics of quantum discord under the effect of
the environmental spin chain has increasingly been investigated~\cite{Liu}. The results show that  quantum discord is more robust than entanglement
for the system
subjected to the environment~\cite{Guo-Jin-Liang:2013aa}.\\

In this letter, we consider two spins coupled to the one-dimensional general quantum
compass model (GQCM) \cite{Eriksson,Nussinov:2009aa,Nussinov:2013aa,Mahdavifar} in the presence of a transverse field~\cite{Jafari1,Jafari2,You:2014aa,Motamedifar-M.:2011aa,You}.
GQCM is a simplified model that describes the nature of the orbital states in the case of a twofold degeneracy \cite{Jafari1}. We analyze the effect of the gapped critical environment on the dynamic evolution of the two-spin
entanglement,  quantum discord, and negativity. Our comprehensive results show  that  induction of the gapped critical environment is nontrivially
different than the gapless critical environment \cite{Quan, Cucchietti3}, and strongly  recommend  it   as a better entangler and quantum channel.
In particular, in the weak coupling regime, QCs at the gapped CP decay rapidly from a maximum and after a very fast initial transient, start oscillating around an average value whereas  show maximum  decay away from   CP.
The results show that at   CP an average value of QCs is enhanced with the increasing of the gap.\\

The full Hamiltonian which considers two spins coupled transversely to the spin $1/2$ general quantum compass chain is fully characterized by % the form
${\cal H}={\cal H}_{E}+{\cal H}_{I}$
with~\cite{Nussinov,Yuan2}
\bea
%\bl
{\cal H}_{E}&=&
\!\!
\sum_{i=1}^{N'}
\Big[
J_{o}
\tilde{\sigma}_{2i-1}^{(+)}
\tilde{\sigma}_{2i}^{(+)}
+J_{e}
\tilde{\sigma}_{2i}^{(-)}
\tilde{\sigma}_{2i+1}^{(-)}
+h(\sigma^{z}_{2i-1}+\sigma^{z}_{2i})
\Big],
\no
\label{eq1}
{\cal H}_{I}&=&
\frac{g}{2}(\sigma^{z}_{A}+\sigma^{z}_{B})
\sum_{i=1}^{N'}(\sigma^{z}_{2i-1}+\sigma^{z}_{2i}),
%\el
\eea
with  ${\cal H}_{I}$ describing the interaction between the central two qubits ($\sigma^{z}_{A},\sigma^{z}_{B}$) and the surrounding
spin chain with  coupling strength $g$.
Additionally,        ${\cal H}_{E}$ is
the Hamiltonian of the environment describing the one-dimensional GQCM (1d-GQCM).
In this representation, 1d-GQCM  is constructed by antiferromagnetic order of
$X$ and $Y$ pseudo-spin components on odd and even bonds at which the pseudo-spin operators are
constructed as   linear combinations of the Pauli matrices $(\sigma^{\alpha=x,y,z})$:
 $\tilde{\sigma}_{2i}^{(\pm)}=\tilde{\sigma}_{i}(\pm\theta)=\cos\theta\sigma^{x}_{i} \pm\sin\theta\sigma^{y}_{i}$
 \cite{You}.
Here $\theta$ ($-\theta$) is arbitrary angle relative to $\sigma^{x}$ for  even (odd) bounds.
$J_{e}$ and $J_{o}$ characterise the   even and odd bound couplings respectively, $h$ is a
transverse field, and $N=2N'$ is the number of spins.

We should emphasise that the 1d-GQCM is exactly solvable with the Jordan-Wigner transformation  \cite{You,Jafari1,Jafari:2015aa}, which in  momentum space leads to
\bea
{\cal H}_E=\sum_{k}\Big[E^{q}_{k}(\gamma_{k}^{q\dag}\gamma_{k}^{q}-\frac{1}{2})+
E^{p}_{k}(\gamma_{k}^{p\dag}\gamma_{k}^{p}-\frac{1}{2})\Big],
\eea
where $\gamma_{k}^{p,q\dag} (\gamma_{k}^{p,q})$ %, and  $\gamma_{k}^{p\dag}, (\gamma_{k}^{p})$
denote independent fermions creation (annihilation) operators. For states with even fermions $E^{q}_{k}=\sqrt{a+\sqrt b}$ and $E^{p}_{k}=\sqrt{a-\sqrt b}$,
 with $a=4h^{2}+|J|^{2}+|L|^{2}$ and
$b=(16h^{2}+2|J|^{2})|L|^{2}+J^{2}{L^\ast}^{2}+{J^\ast}^{2}L^{2}$,
where  the  parameters $L$ and $J$ are defined  by
$L=(J_o+J_ee^{\oi k})/4$, and
$J=(J_o e^{\oi \theta} - J_e e^{\oi (k-\theta)})/4$. % and the ground state energy is given by $E_{G}=-\frac{1}{2}\sum_{k}(E^{q}_{k}+E^{p}_{k})$.

It is known that the ground state is separated from the lowest-energy pseudo-spin excitation by a pseudo-spin gap which vanishes at $\cos\theta_{c}=h/\sqrt{J_{o}J_{e}}$~\cite{You}.
First we concentrate on an idiosyncratic case of $\theta_c=\pi/2$ in the absence of external magnetic field.
For this case,  the ground-state has a macroscopic degeneracy of $2^{N/2-1}$ away from the isotropic point ($J_{o}\neq J_{e}$), which becomes $2^{N/2}$ when the orbital interactions are isotropic. The gap at $k=\pi$ is given by the anisotropy of the pseudo-spin exchange, $\Delta=|J_{e}-J_{o}|$, and just vanishes at $J_{e}=J_{o}$
which implies that the degeneracy increases by an additional factor of 2 due to the band-edge points.
It has been proven that in the absence of an external magnetic field, for  $\theta_c=\pi/2$,  GQCM with $Z_2$ symmetry is critical for arbitrary $J_{e}/J_{o}$~\cite{You, Nussinov}.
QPT takes place between two different disordered phases where the
 model exhibits highest possible frustration of interactions~\cite{You, Nussinov}.
 The isotropic point corresponds to a multicritical point where the gap closes
quadratically at $k=\pm\pi$ as a result of the confluence of two Dirac points~\cite{Niu}.

Since the central two qubits operators ($\sigma^{z}_{A},\sigma^{z}_{B}$) and
 the environmental spin chain ($\sigma^{\alpha}_{i}$) satisfy the commutation relation $[\sigma^{z}_{A}+\sigma^{z}_{B},\sigma^{\alpha}_{i}]=0$,  the total Hamiltonian %Eq. (\ref{eq1})
  can be rewritten as
${\cal H}=\sum_{\mu=1}^{4}|\varphi_{\mu}\rangle\langle\varphi_{\mu}|\otimes {\cal H}^{h_{\mu}}_{E}$.
 $|\varphi_{\mu}\rangle,~(\mu=1, \ldots,4)$ denotes
 the $\mu$th eigenstate of the operator $\frac{g}{2}(\sigma^{z}_{A}+\sigma^{z}_{B})$ corresponding to the $\mu$th eigenvalue
 $\varepsilon_{\mu}$ ($\varepsilon_{1}=g,~\varepsilon_{2}=\varepsilon_{3}=0,~\varepsilon_{4}=-g$).
 Here ${\cal H}^{h_{\mu}}_{E}$ is defined through  ${\cal H}_{E}$ by replacing   $h$ with $h_{\mu}$,
 where  $h_{\mu}=h+\varepsilon_{\mu}$.
We suppose that  the initial state of the total system is disentangled with
$\rho_{tot}(0)=\rho_{AB}(0)\otimes\rho_{E}(0)$.
Considering
 $|\psi_{E}\rangle$ as an initial state of the environmental spin
chains,
and $\rho_{AB}(0)$ and $\rho_{E}(0)=|\psi_{E}\rangle\langle\psi_{E}|$ as initial density matrix state of the two-qubits system and   environment respectively,
%.
 the evolution of the total system will be governed by $\rho_{tot}(t)= U(t)\rho_{tot}(0)U^{\dag}(t)$.
 Accordingly, the reduced density matrix of two-qubits $AB$ is obtained by tracing out the environment~\cite{Yuan2},
\bea
\label{eq2}
\rho_{AB}(t)=\sum_{\mu,\nu=1}^{4}F_{\nu\mu}(t)\langle\varphi_{\nu}|\rho_{AB}(0)|\varphi_{\mu}\rangle
|\varphi_{\nu}\rangle\langle\varphi_{\mu}|,
\eea
where the decoherence factors are achieved by $F_{\nu\mu}=\langle\psi_{E}|U^{ h_{\mu}\dag}_{E}  U^{h_{\nu}}_{E}|\psi_{E}\rangle$, and $U^{h_{\nu}}_{E}=U^{h_{\nu}}_{E}(t)=e^{-\oi {\cal H}^{h_{\nu}}_{E}t}$ is the time evolution operator driven by ${\cal H}^{h_{\nu}}_{E}$.
By considering the ground state of the environment spin chain as an initial state, the decoherence factors reduce to a LE form given in Ref.~\cite{Quan}, which is a dynamical version of the ground-state fidelity \cite{Montes,Jafari:2015aa}.
We assume that the two qubits $AB$ initially stem from the $X$-structure states $\rho_{AB}(0)=\frac{1}{4}(I_{AB}+\sum_{\alpha}c_{\alpha}\sigma^{\alpha}_{A}\otimes\sigma^{\alpha}_{B})$
with  the identity operator on two-qubits system, $I_{AB}~$\cite{Englert:2001aa}. The parameters $c_{\alpha=x,y,z}$ are chosen to be real %parameters
 that insure that  $\rho_{AB}(0)$ is a legal quantum state.
This state
is chosen in a general form  to contain Bell-diagonal states and Werner states.
According to Eq. (\ref{eq2}), the reduced density matrix in the standard
basis
($\upup,\; \updown,\; \downup,\; \downdown$) can be written as
%${|\downarrow\downarrow\rangle, |\downarrow\uparrow\rangle, |\uparrow\downarrow\rangle, |\uparrow\uparrow\rangle}$
%
\be
\bl
\label{eq3}
\rho_{AB}(t)=\frac{1}{4}\left(
               \begin{array}{cccc}
                 1+c_{z} & 0 & 0 & c_{\beta} \\
                 0 & 1-c_{z} & c_{\gamma} & 0 \\
                 0 & c_{\gamma}^{\ast} & 1-c_{z} & 0 \\
                 c_{\beta}^{\ast} & 0 & 0 & 1+c_{z} \\
               \end{array}
             \right),
\el
\ee
with $c_{\beta}=(c_{x}-c_{y})F_{14}$, and $c_{\gamma}=(c_{x}+c_{y})F_{23}$.

%%%%%%%%%%%%%%%%%%%%%%%  Fig.2   %%%%%%%%%%%%%%%%%%%%%%%
\begin{figure}
\includegraphics[width=\linewidth]{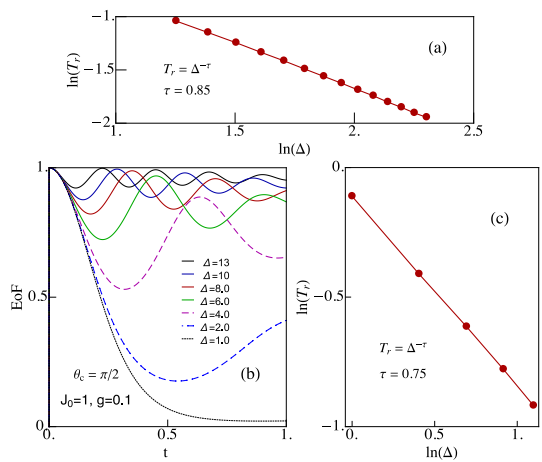}
\caption{(Color online)
(a) The scaling behavior of the relaxation time $T_{r}$ in terms of
energy gap for $N=400, J_{o}=1, J_{e}=4$ and $g=0.1$.
 (b) Evaluation of entanglement as a function of time for different energy
gap values.
(c) The scaling behaviour of relaxation for small values of gap.
}
\label{fig2}
\end{figure}
%%%%%%%%%%%%%%%%%%%%%%%%%%%%%%%%%%%%%%%%%%%%%%%%%%%%%%%%

To quantify the entanglement dynamics of two qubits $AB$, we utilize the concurrence directly to calculate
the entanglement of formation (EoF). The entanglement is a monotonically increasing function of concurrence ($C'$), which is defined by~\cite{Wootters:1998aa}
$${\rm EoF}=-f_{C'}\log_{2}f_{C'}-(1-f_{C'})\log_{2}(1-f_{C'}),$$
with $f_{C'}=\frac{1}{2}(1+\sqrt{1-C'^{2}})$
and $C'=\max\{\lambda_{1}-\lambda_{2}-\lambda_{3}-\lambda_{4},0\}$, where $\lambda_{i}$ %, $(i=1,\ldots,4)$
are the square roots of the eigenvalues in descending order of the operator $\rho_{AB}(t)(\sigma^{y}_{A}\otimes\sigma^{y}_{B})\rho^{\ast}_{AB}(t)(\sigma^{y}_{A}\otimes\sigma^{y}_{B})$.
One can directly calculate the concurrence for the state defined  by Eq.~(\ref{eq3})
as
$C'=\max\{\frac{|c_{\beta}|+c_{z}-1}{2},\frac{|c_{\gamma}|-c_{z}-1}{2},0\}$.
Moreover, quantum discord is given by
\be
\bl
&
Q_{AB}=
\frac{1}{4}\sum_{\pm}
\Big[
(1-c_{z} \pm |c_{\gamma}|)\log_{2}(1-c_{z}\pm |c_{\gamma}|)+
\\
&\hspace{0.3cm}
(1+c_{z} \pm |c_{\beta}|)\log_{2}(1+c_{z}  \pm |c_{\beta}|)
\Big]
-C
\Big(
\rho_{AB}(t)
\Big),
\el
\ee
where
$C(
\rho_{AB}
)=\sum_{\pm} \frac{1\pm \vartheta}{2}\log_{2}(1\pm \vartheta)$,
is the classical correlation
%and is defined by
with $\vartheta=\max\{|c_{z}|, \frac{|c_{\beta}|+|c_{\gamma}|}{2}\}$.
To investigate the  time evolution of QCs, we set the parameters $c_{x,z}=-c_{y}=1$, so that  the initial state becomes the Bell state
$(\downdown+\upup)/\sqrt{2}$.
For this case, concurrence and quantum discord become $C'=|F_{14}(t)|$ and
 $Q=\sum_{\pm}
 \frac{1\pm |F_{14}(t)|}{2}\log_{2}(1 \pm |F_{14}(t)|)
 $,
  respectively.
  To illustrate the dynamical properties of QCs, we carry out the numerical calculation using the exact expression.
The density plots of the  time evolution of  entanglement and quantum discord have been depicted
in the weak coupling limit in Fig.~\ref{fig1}(a \& b).
We set the Hamiltonian parameters in such a way that
 the system becomes  gapped at CP $\theta_c=\pi/2$.
 Obviously both entanglement and quantum discord are   monotonically increasing functions
of the decoherence factor $|F_{14}(t)|$.
Thus, quantum discord behaves in a similar
way as entanglement does and is always less than entanglement in the process of evolution.

In contrast with the above situation, Fig.~\ref{fig1}(c)  represents a  density plot of entanglement for the gapless critical case.
As is clear, the enhanced decay of   QCs induced by quantum criticality of the surrounding environment is broken by gapped quantum critical environment and the maximum decaying happens away from  CP.
The result shows that in the weak coupling regime, QCs decay from their maximum and after an initial transient, start oscillating around an average  value.
As observed in Fig.~\ref{fig1}, % (a), (b) and (c),
 QCs are symmetric with respect to $\theta=\pi/2$ and the numerical calculation shows that the valley narrows as $g$ decreases and system size, $N$, increases.
A more detailed analysis also shows that the relaxation time $T_{r}$ at which  QCs decay to their local minimums at  CP of the environment (see Fig.~\ref{fig2}(b)), reveals a power-law singularity as a function of the gap $\Delta$. This is presented in Fig.~\ref{fig2}(a) which specifies a linear behaviour of $\ln(T_{r})$ versus $\ln(\Delta)$. The scaling behaviour is obtained as $T_{r}=|\Delta|^{-1/\tau}$ with the exponent $\tau=-0.75$ for small value of the gap whereas the exponent $\tau$ equals $-0.85$ for large values of the gap. Thus, one can conclude
that the appearance of the two energy scales in the system is appropriate with the
number of  states which are  involved. It means that the dynamics of QCs at very large energy gap values  mainly originates from the ground state and the excited states have a very tiny contribution.
%% to the dynamic of QCs.
It would be worth mentioning  that the interaction coupling $g$ does not affect the exponent $\tau$ in the weak coupling limit, as displayed  by numerical simulations.

\begin{figure}[!t]
\centerline{\includegraphics[width=\linewidth]{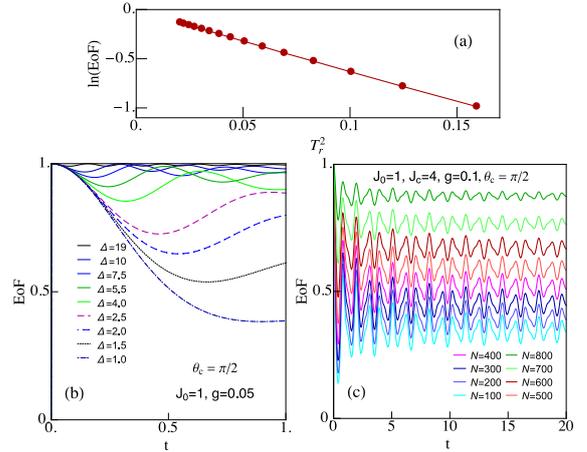}}
\caption{(Color online)
(a) Scaling of the minimum of quantum correlations
at  critical point $(\theta_c=\pi/2)$ for systems of various gaps (Fig.~\ref{fig2}b).
(b)
Entanglement evaluation as a function of time for different energy
gap for $N=400$, $J_{o}=1$, and $ g=0.05$.
(c) Entanglement as a function of time at  critical point
$(\theta_c=\pi/2)$ for various system sizes, and  $J_{o}=1$, $J_{e}=4$, and $g=0.1$.
%.
 }
\label{fig3}
\end{figure}

To study the scaling behaviour of QCs at  CP, we have derived the scaling behaviour of local minimum values of QCs versus the second power of the relaxation time.
This has been plotted in Fig.~\ref{fig3}(a), which shows the linear behaviour of $\ln [{\rm EoF}(T_{r})]$, the same as   $\ln [Q(T_{r})]$, versus $T^{2}_{r}$.
In other words, the local minimum values of QCs scale exponentially with the second power of the relaxation time, ${\rm EoF}(T_{r})=\exp(-\delta T^{2}_{r})$ with exponent $\delta\propto(-\Delta/g)$.
In particular, our results imply that the decay of  the local minimum of QCs at  CP of environment enhances with   decreasing energy gap and increasing  interaction coupling (see Fig.~\ref{fig3}(b)).

However, oscillations of QCs around an average  value increases as the energy gap of the environmental spin chain increases.
On the other hand,  Fig.~\ref{fig3}(c) reveals an interesting phenomenon in  QCs at  CP ($\theta_c=\pi/2$).
 The periods of the revival of  QCs is independent of the size of the environment which means that   CP is a scale invariance point where  quantum fluctuations extend over all length scales.
  Moreover, a similar result also has  confirmed the above argument for   the time evolution of QCs from mixed states  $c_{x}=1$, and $c_{y}=-c_{z}\in [0,1]$.
The  presented results presented here indicate that in the strong coupling regime decoherence factors and QCs decay to zero  in a very short time as opposed to what happens in the weak coupling regime.
As the coupling strength increases, the valleys widen and the influence of the energy gap on the generation of QCs decreases even when
approached along the gapped critical point.\\

%%%%%%%%%%%%%%%%%%%%%%%  Fig.3  %%%%%%%%%%%%%%%%%%%%%%%
%%%%%%%%%%%%%%%%%%%%%%%%%%%%%%%%%%%%%%%%%%%%%%%%%%%%%%%

%
%
In the presence of a magnetic field, GQCM shows a gapless phase transition  and our results support previous studies~\cite{Yuan2,Cucchietti3,Mostame,Sun:2007aa,Cheng:2009aa}.
We also examine  two qutrits coupled to one 1d-GQCM~\cite{Sun:2007ab}. The negativity displays almost the same dynamical behaviour as entanglement and quantum discord do.\\

In summary, using the  general quantum compass model as an environmental system, the dynamical evolution of the decoherence factors, quantum correlations, and negativity  of the central spins has been
% numerically and analytically
investigated for different initial states. The relation between the quantum-classical transition of the central system, and the occurrence of an avoided level crossing quantum phase transition in its surrounding system has been analysed. It is well known that  the gapless quantum criticality enhances decaying of decoherence factors,
while our calculations  represent a different story for gapped critical environment. The finding results show that   long-time quantum correlations at the critical point is an  effect of gapped criticality, and maximum decaying occurs away from the critical point.

The role of the gapped critical spin chain is to prevent the complete drain of information from central systems to the environment~\cite{Aguilar:2014aa} and provides them a better environment for preserving quantum correlations.
In other words, the amount of decoherence which  travels into the central spin state depends on the excited states of the  environment. Hence the energy gap could block the propagation of decoherence along the environment and consequently reduces its effect on the central spin.
These results highlight the current outlook of using quantum spin chains as entanglers or quantum channels in quantum information devises~\cite{Rio:2011aa,Sahling:2015aa}.
Besides, quantum gapped criticality  may have potential applications in quantum computations.
\\

%%%%%%%%%%%%%%%%%%%%%%%%%%%%%%%%%%%%%%%%%%%%%%%%%%%%%%%%%%%%%%%%%%%%%%%%%%%%%%%%%%%%%%%%%
{\it Acknowledgments}:
We would like to thank S. Mahdavifar for useful discussions.   A.A.  acknowledges %partial  financial
support by  Max Planck POSTECH/KOREA Research Initiative (No. 2011-0031558) programs through NRF funded by
MSIP of Korea.

%%%%%%%%%%%%%%%%%%%%%%%%%%%%%%%%%%%%%%%%%%%%%%%%%%%%%%%%%%%%%%%%%%%%%%%%%%%%%%%%%%%%%%%%%
%%%%%%%%%%%%%%%%%%%%%%%%%%%%%%%%%%%%%%%%%%%%%%%%%%%%%%%%%%%%%%%%%%%%%%%%%%%%%%%%%%%%%%%%%
%%%%%%%%%%%%%%%%%%%%%%%%%%%%%%%%%%%%%%%%%%%%%%%%%%%%%%%%%%%%%%%%%%%%%%%%%%%%%%%%%%%%%%%%%
%
%{References}
%
\bibliographystyle{EPL}
\bibliography{Ref}
%%%%%%%%%%%%%%%%%%%%%%%%%%%%%%%%%%%%%%%%%%%%%%%%%%%%%%%%%%%%%%%%%%%%%%%%%%%%%%%%%%%%%%%%%
\end{document}